\pdfoutput=1
\documentclass[aps,prl,a4paper
               ,amsmath,amssymb
               ,twocolumn,showpacs
               ,floatfix
               ]{revtex4-1}

\usepackage{dcolumn}

\usepackage{amsfonts}
\usepackage{tikz}
\usepackage{tabularx} \newcolumntype{C}{>{\centering}X}
\usepackage{graphicx}
\usepackage{hyperref}
\usepackage{wrapfig}
\usepackage{ragged2e}
\usepackage{braket}
\usepackage{bm}
\usepackage{amsmath}
\usepackage[caption=false]{subfig}

\def\tp{\otimes}
\def\ds{\oplus}
\def\Z{\mathbb{Z}}

\def\L{\mathcal{L}}
\def\Wop[#1][#2][#3][#4][#5]{W\left(\left.\begin{array}{cc} #1&#2\\#3&#4\end{array}\right|#5\right)}
\def\WopAlt[#1][#2][#3][#4][#5][#6][#7]{W^{#6#7}\left(\left.\begin{array}{cc} #1&#2\\#3&#4\end{array}\right|#5\right)}

\begin{document}
\title{Quantum phases of a chain of strongly interacting anyons}

\author{Peter E.~Finch}
\author{Holger Frahm}
\author{Marius Lewerenz}
\author{Ashley Milsted}
\author{Tobias J.~Osborne}

\affiliation{%
Institut f\"ur Theoretische Physik, Leibniz Universit\"at Hannover,
Appelstra\ss{}e 2, 30167 Hannover, Germany}


\begin{abstract}
  Quantum gates for the manipulation of topological qubits rely on
  interactions between non-Abelian anyonic quasiparticles. We study the
  collective behaviour of systems of anyons arising from such interactions. In
  particular, we study the effect of favouring different fusion channels of
  the screened Majorana spins appearing in the recently proposed topological
  Kondo effect. Based on the numerical solution of a chain of $SO(5)_2$ anyons
  we identify two critical phases whose low-energy behaviour is characterised
  by conformal field theories with central charges $c=1$ and $c=8/7$,
  respectively.  Our results are complemented by exact results for special
  values of the coupling constants which provide additional information about
  the corresponding phase transitions.
\end{abstract}

\maketitle

%
%

Low-dimensional quantum systems hold an irresistible and enduring fascination
because they can support topological states of matter with exotic
quasiparticles, \emph{anyons}, exhibiting unusual braiding statistics
\cite{wilczek:1990a}.  While initially a curiosity, anyons
generated considerable excitement when it was realized that the fractional
quantum Hall effect \cite{Moore1991, *Read1999} --- and later nanowires
\cite{Stanescu2011, MZFP12} and the $p_x+ip_y$ superconductor \cite{Read2000}
--- support these fractionalized excitations.  Further interest was
recently sparked by the remarkable proposal that systems of non-Abelian anyons could carry out fault-tolerant quantum computation via braiding 
\cite{dennis:2002a, nayak:2008a, kitaev:2003a, kitaev:2006a,wang:2008a,
  pachos:2012a, barends:2014a}.

An increasing number of scenarios for the realization of localized anyon
modes in electronic materials have been proposed \cite{beenakker:2013a} leading to the recent experimental observation of signatures of Majorana fermions (or Ising anyons) at the ends of nanowires coupled to a superconductor \cite{Stanescu2011, MZFP12}.  
However, braiding of Ising anyons alone does not yield a set of gates sufficient for
universal quantum computation: topological qubits are found by additionally coupling $M$ of these quasiparticles to form an
$SO(M)$ Majorana spin. 

When connected to electronic leads via tunnel junctions Majorana topological
spins exhibit the recently proposed ``topological'' Kondo effect displaying
strong non-Fermi liquid correlations \cite{BeCo12}.  Exploiting a combination
of both perturbative renormalization group analysis and conformal field theory
methods the corresponding Kondo fixed point has been identified with an
$SO(M)_2$ Wess-Zumino-Novikov-Witten boundary conformal field theory
\cite{BeCo12,ABET13}.
A proposed scheme to probe and manipulate these Kondo-screened Majorana spins
\cite{ABET13,ABET14} constitutes a first step towards a realization of quantum
gates. Indeed it was shown that $SO(3)_2$ anyons (and likely $SO(p)_2$ for prime
$p\ge 5$) are able to support universal quantum computation\cite{cui:2014a}.
Thus the study of models for $SO(M)_2$ anyons has become an intriguing and
relevant possibility and introducing couplings between such objects is now a
natural next step.


Unfortunately systems of anyons are extremely difficult to study: one must
keep track of their entire space-time history in order to discuss their
dynamics. While the non-interacting case is now becoming well understood (see,
e.g., \cite{wang:2008a}) the classification of phases for systems of
\emph{interacting} anyons has progressed much slower. An additional
complication is that the description of the dynamics of a highly entangled
$SO(M)$ Majorana spin in the topological Kondo model, and the collective
behaviour of many such subsystems, involves dealing with strong correlations.
Indeed, only recently have \emph{one-dimensional} interacting systems, where
the anyon model may be seen as a deformation of $SU(2)$, been studied in
earnest.
Using exact numerical diagonalization, partly complemented by analytical
results known for the related restricted solid on solid (RSOS) models
\cite{AnBF84,Kakashvili2012}, it has been possible to identify the possible
phases realized in these models \cite{feiguin:2007a, GATH13}.  These models
are not just toys: the study of one-dimensional systems offers nontrivial
insight into the more general case as they mediate the boundary between
different topological phases \cite{Gils2009}.

In one dimension we can exploit powerful tensor-network variational methods falling
under the umbrella of the \emph{density matrix renormalization group} (DMRG)
\cite{schollwoeck:2005a,schollwock:2011a}. These methods, with
impetus from the study of \emph{quantum entanglement}, have led to unparalleled insights in recent years overcoming many previously insurmountable roadblocks, including,
the simulation of dynamics \cite{vidal:2003a, haegeman:2011b} and fermions
\cite{corboz:2009a, corboz:2010a, corboz:2010b, kraus:2010a} without sign
problems, the determination of spectral information
\cite{haegeman:2012a}, and higher dimensions by exploiting tensor networks including
the projected entangled-pair states (PEPS) \cite{verstraete:2004a} and the
multiscale entanglement renormalization ansatz (MERA) \cite{vidal:2006a,
vidal:2007a}. This progress has been recently extended to the anyonic setting with the development of anyonic matrix product states (MPS) and MERA 
routines to obtain ground-state properties \cite{koenig:2010a, pfeifer:2010a} 
and to simulate the spectrum and dynamics of interacting anyons \cite{singh:2013a,poilblanc:2013a, zatloukal:2012a, zaletel:2013a},
 providing an attractive complement to other methods such as
exact diagonalization \cite{trebst:2008a, Gils2009, poilblanc:2011a,
  poilblanc:2012a} and Monte Carlo \cite{tran:2010a}.

The purpose of this letter is to exploit both analytical and numerical tools to carry out a comprehensive study of a nontrivial system of anyons relevant to the topological Kondo effect. Specifically, as an effective model, we consider the fundamental quasi-particles in an $SO(5)_2$
Chern-Simons theory \cite{hastings:2013a}.  Using two complementary
techniques, one based on the Bethe ansatz and the other on cutting-edge
tangent-plane tensor network methods, we investigate the ground-state and low-lying excitations of a one-dimensional condensate of these non-Abelian anyons and characterize and classify their critical phases and integrable points.

\begingroup
\squeezetable
\begin{table}[ht]
\caption{\label{tabfus}Fusion rules for the $SO(5)_{2}$ anyons}
\begin{tabular}{|c|c|c|c|c|c|c|} \hline
	$\tp$ & $\psi_{1}$ & $\psi_{2}$ & $\psi_{3}$ & $\psi_{4}$ & $\psi_{5}$ & $\psi_{6}$ \\ \hline
	$\psi_{1}$& $\psi_{1}$ & $\psi_{2}$ & $\psi_{3}$ & $\psi_{4}$ & $\psi_{5}$ & $\psi_{6}$ \\ \hline
	$\psi_{2}$& $\psi_{2}$ & $\psi_{1}\ds\psi_{5}\ds\psi_{6}$ & $\psi_{3}\ds\psi_{4}$ & $\psi_{3}\ds\psi_{4}$ & $\psi_{2}\ds\psi_{5}$ & $\psi_{2}$ \\ \hline
	$\psi_{3}$& $\psi_{3}$ & $\psi_{3}\ds\psi_{4}$ & $\psi_{1}\ds\psi_{2}\ds\psi_{5}$ & $\psi_{2}\ds\psi_{5}\ds\psi_{6}$ & $\psi_{3}\ds\psi_{4}$ & $\psi_{4}$ \\ \hline
	$\psi_{4}$& $\psi_{4}$ & $\psi_{3}\ds\psi_{4}$ & $\psi_{2}\ds\psi_{5}\ds\psi_{6}$ & $\psi_{1}\ds\psi_{2}\ds\psi_{5}$ & $\psi_{3}\ds\psi_{4}$ & $\psi_{3}$ \\ \hline
	$\psi_{5}$& $\psi_{5}$ & $\psi_{2}\ds\psi_{5}$ & $\psi_{3}\ds\psi_{4}$ & $\psi_{3}\ds\psi_{4}$ & $\psi_{1}\ds\psi_{2}\ds\psi_{6}$ & $\psi_{5}$ \\ \hline
	$\psi_{6}$& $\psi_{6}$ & $\psi_{2}$ & $\psi_{4}$ & $\psi_{3}$ & $\psi_{5}$ & $\psi_{1}$ \\ \hline
\end{tabular}
\end{table}
\endgroup
\noindent

The fusion rules for $SO(5)_2$, the truncation of the category of irreducible
representations of the quantum group $U_{q}(so(5))$, with
$q=e^{\frac{2i\pi}{5}}$ \cite{Bond07}, are given in Table~\ref{tabfus}.  They
are diagonalized by the modular $S$-matrix
\begin{equation}
\label{smat}
  S  = \frac{1}{2\sqrt{5}}
  \left(\begin{array}{cccccc}
         1 & 2 & \sqrt{5} & \sqrt{5} & 2 & 1 \\
         2 & -2\phi & 0 & 0 & 2\phi^{-1} & 2 \\
         \sqrt{5} & 0 & -\sqrt{5} & \sqrt{5} & 0 & -\sqrt{5} \\
         \sqrt{5} & 0 & \sqrt{5} & -\sqrt{5} & 0 & -\sqrt{5} \\
         2 & 2\phi^{-1} & 0 & 0 & -2\phi & 2 \\
         1 & 2 & -\sqrt{5} & -\sqrt{5} & 2 & 1 \\
        \end{array} \right),
\end{equation}
where $\phi = \frac{1+\sqrt{5}}{2}$. Specifically, we consider fusion paths of
length $\mathcal{L}$ for $\psi_3$ anyons, represented diagrammatically as
\begin{center} \begin{small} \begin{tikzpicture}[scale=0.8]
	\tikzstyle{every node}=[minimum size=0pt,inner sep=0pt]
	\tikzstyle{every loop}=[]
	\node (ns) at (0.2,0.0) {$\psi_{a_{0}}$};
	\node (nm) at (4.0,0.0) {\phantom{m}\ldots\phantom{m}};
	\node (ne) at (6.8,0.0) {$\psi_{a_{\L}}$};
	\node (n1t) at (1.0,0.7) {$\psi_{3}$};
	\node (n2t) at (2.0,0.7) {$\psi_{3}$};
	\node (n3t) at (3.0,0.7) {$\psi_{3}$};
	\node (n4t) at (5.0,0.7) {$\psi_{3}$};
	\node (n5t) at (6.0,0.7) {$\psi_{3}$};
	\node (n1b) at (1.0,0.0) {};
	\node (n2b) at (2.0,0.0) {};
	\node (n3b) at (3.0,0.0) {};
	\node (n4b) at (5.0,0.0) {};
	\node (n5b) at (6.0,0.0) {};
	\node (l1) at (1.6,-0.3) {$\psi_{a_{1}}$};
	\node (l2) at (2.6,-0.3) {$\psi_{a_{2}}$};
	\node (l3) at (3.6,-0.3) {$\psi_{a_{3}}$};
	\node (l4) at (5.6,-0.3) {$\psi_{a_{\L-1}}$};
	\foreach \from/\to in {ns/nm,nm/ne} \draw (\from) -- (\to);
	\foreach \from/\to in {n1t/n1b,n2t/n2b,n3t/n3b,n4t/n4b,n5t/n5b} \draw
        (\from) -- (\to); 
      \end{tikzpicture} \end{small} 
\end{center} 
where fusing occurs from top-left to bottom-right.
Below these paths are identified with basis vectors of an anyonic Hilbert
space, i.e.\ $\Ket{a_{0}a_{1}\cdots a_{\mathcal{L}}}$, where neighbouring labels must be
related through fusion with $\psi_3$. This is equivalent to labels $a_i$ and
$a_{i+1}$ being adjacent on following graph
\begin{center} \begin{small} 
\begin{tikzpicture}[scale=0.6]
	\tikzstyle{every node}=[circle,draw,thin,fill=blue!40,minimum
        size=12pt,inner sep=0pt] 
	\tikzstyle{every loop}=[]
	\node (n0) at (0.0,0.0) {$1$};
	\node (n1) at (3.0,1.0) {$2$};
	\node (n2) at (1.5,0.0) {$3$};
	\node (n3) at (4.5,0.0) {$4$};
	\node (n4) at (3.0,-1.0) {$5$};
	\node (n5) at (6.0,0.0) {$6$};
	\foreach \from/\to in {n0/n2,n1/n2,n1/n3,n2/n4,n3/n4,n3/n5} \draw (\from) -- (\to);
      \end{tikzpicture} \end{small}
\end{center}

Ordering of the fusion can be changed by means of $F$-moves (i.e.\ generalized
6-$j$ symbols),
\begin{center} \begin{small} \begin{tikzpicture}[scale=0.8]
	\tikzstyle{every node}=[minimum size=0pt,inner sep=0pt]
	\tikzstyle{every loop}=[]
	\node (ns) at (0.2,0.0) {$\psi_{a_{i-1}}$};
	\node (ne) at (2.8,0.0) {$\psi_{a_{i+1}}$};
	\node (n1t) at (1.0,0.7) {$\psi_{3}$};
	\node (n2t) at (2.0,0.7) {$\psi_{3}$};
	\node (n1b) at (1.0,0.0) {};
	\node (n2b) at (2.0,0.0) {};
	\node (l1) at (1.6,-0.3) {$\psi_{a_{i}}$};
	\foreach \from/\to in {ns/ne} \draw (\from) -- (\to);
	\foreach \from/\to in {n1t/n1b,n2t/n2b} \draw (\from) -- (\to);
	\node (l2) at (4.8,0.0) {$=\sum_{a_{i}'} (F^{a_{i-1}33}_{a_{i+1}})^{a_{i}}_{a_{i}'}$}; 
	\node (vs) at (6.9,-0.2) {$\psi_{a_{i-1}}$};
	\node (ve) at (9.5,-0.2) {$\psi_{a_{i+1}}$};
	\node (v1) at (7.7,0.7) {$\psi_{3}$};
	\node (v2) at (8.7,0.7) {$\psi_{3}$};
	\node (v12) at (8.2,0.2) {};
	\node (vb) at (8.2,-0.2) {};
	\node (l3) at (8.55,0.05) {$\psi_{a_{i}'}$};
	\foreach \from/\to in {vs/ve} \draw (\from) -- (\to);
	\foreach \from/\to in {v1/v12,v2/v12,v12/vb} \draw (\from) -- (\to);
\end{tikzpicture} \end{small}
\end{center}
For the $SO(5)_{2}$ fusion rules there exist four known sets of inequivalent
unitary $F$-moves, each of which corresponds to a different $S$-matrix
\cite{Bond07}.  Hence our choice (\ref{smat}) of the $S$-matrix determines the
$F$-moves which can be used to construct two-site projection operators,
\begin{equation}
\begin{aligned}
  p^{(b)}_{i}
  & = \sum_{a_{i-1},a_{i},a_{i}',a_{i+1}} \left[\left(F^{a_{i-1}jj}_{a_{i+1}}\right)^{a_{i}'}_{b}\right]^{*} \left(F^{a_{i-1}jj}_{a_{i+1}}\right)^{a_{i}}_{b} \\
  & \Ket{\cdots a_{i-1}a_{i}'a_{i+1}\cdots}\Bra{\cdots  a_{i-1}a_{i}a_{i+1}\cdots}.
\end{aligned} \label{eqnproj}
\end{equation}
We can couple pairs of 
$\textsl{SO}(5)_2$ anyons via these projection operators leading to a chain of
$SO(5)_2$ anyons with nearest-neighbour interactions subject to periodic
boundary conditions $a_0\equiv a_{\mathcal{L}}$:
\begin{equation}
\label{eqhamil}
  \mathcal{H}_{\theta}  =  \sum_{i=1}^{\L} 
  \left( \cos\left(\frac{\pi}{4}+\theta\right) p^{(2)}_{i} +
  \sin\left(\frac{\pi}{4}+\theta\right) p^{(5)}_{i} \right)\,.
\end{equation}
From the fusion rules we see that there exists an automorphism exchanging
$\psi_2$ and $\psi_5$ which allows one to construct a non-local unitary
transformation mapping $\mathcal{H}_{\theta} \leftrightarrow
\mathcal{H}_{-\theta}$.

An important property of anyonic models of this type is the presence of
topological charges that commute with each other and the global Hamiltonian:
\begin{equation}
\label{topoy}
  \Bra{a_{1}'\cdots a_{\mathcal{L}}'} Y_{\ell} \Ket{a_{1}\cdots a_{\mathcal{L}}}  =
  \prod_{i=1}^{\mathcal{L}} \left(F^{\ell a_{i}'3}_{a_{i+1}}\right)^{a_{i}}_{a_{i+1}'}\,.
\end{equation}
The action of these operators corresponds to the insertion of an auxiliary anyon
$\psi_\ell$, which is then moved around the ring by application of $F$-moves,
and finally removed again \cite{kitaev:2006a,GATH13}.
The eigenvalues of $Y_{\ell}$ are given in terms of the elements of the
$S$-matrix (\ref{smat}) as $\frac{S_{j\ell}}{S_{1\ell}}$ \cite{kitaev:2006a}.

%
To study the model (\ref{eqhamil}) we have employed a combination of numerical
and analytical techniques. We exploit the symmetry under exchange of $\psi_2$ and
$\psi_5$ mentioned above as an additional check for our numerical results.

\paragraph{Numerical analysis.}

We simulated the $\psi_3$ anyon chain numerically using the evoMPS
software package \cite{EvoMPS} via the time-dependent
variational principle (TDVP) in imaginary time to calculate an
approximate matrix product state representation of the ground state 
 \cite{Werner1992, Verstraete2008}. The TDVP approximately
solves the Schr\"odinger equation for an infinitesimal time step,
providing flow equations for the MPS variational parameters \cite{haegeman:2011b}. In imaginary time the convergence of these flow equations can be enhanced via the nonlinear conjugate gradient method \cite{milsted_phi4}.

For this work we used uniform MPS to represent translation-invariant 
states of the chain in the thermodynamic limit. (Note that here a
``site'' in the chain corresponds to a label in the fusion path, not a
site where a $\psi_3$-anyon sits.)  Since the 
system is invariant under translations of two sites, we blocked
 adjacent labels into a virtual site. We also
exploit a symmetry which allows us to only consider states with labels
$1,2,5,6$ on odd sites and labels $3,4$ on even sites. This
simplification results in a local dimension of $6$ for the virtual
sites.

A state constructed in this way is still not generally a state of the anyon
chain, because it is only guaranteed to be compatible with the fusion rules
within a single virtual site: we added penalty
terms to the Hamiltonian to suppress states not corresponding to a valid
fusion path.  

\begin{figure}[htp]
    \centering
    \includegraphics[width=.95\linewidth]{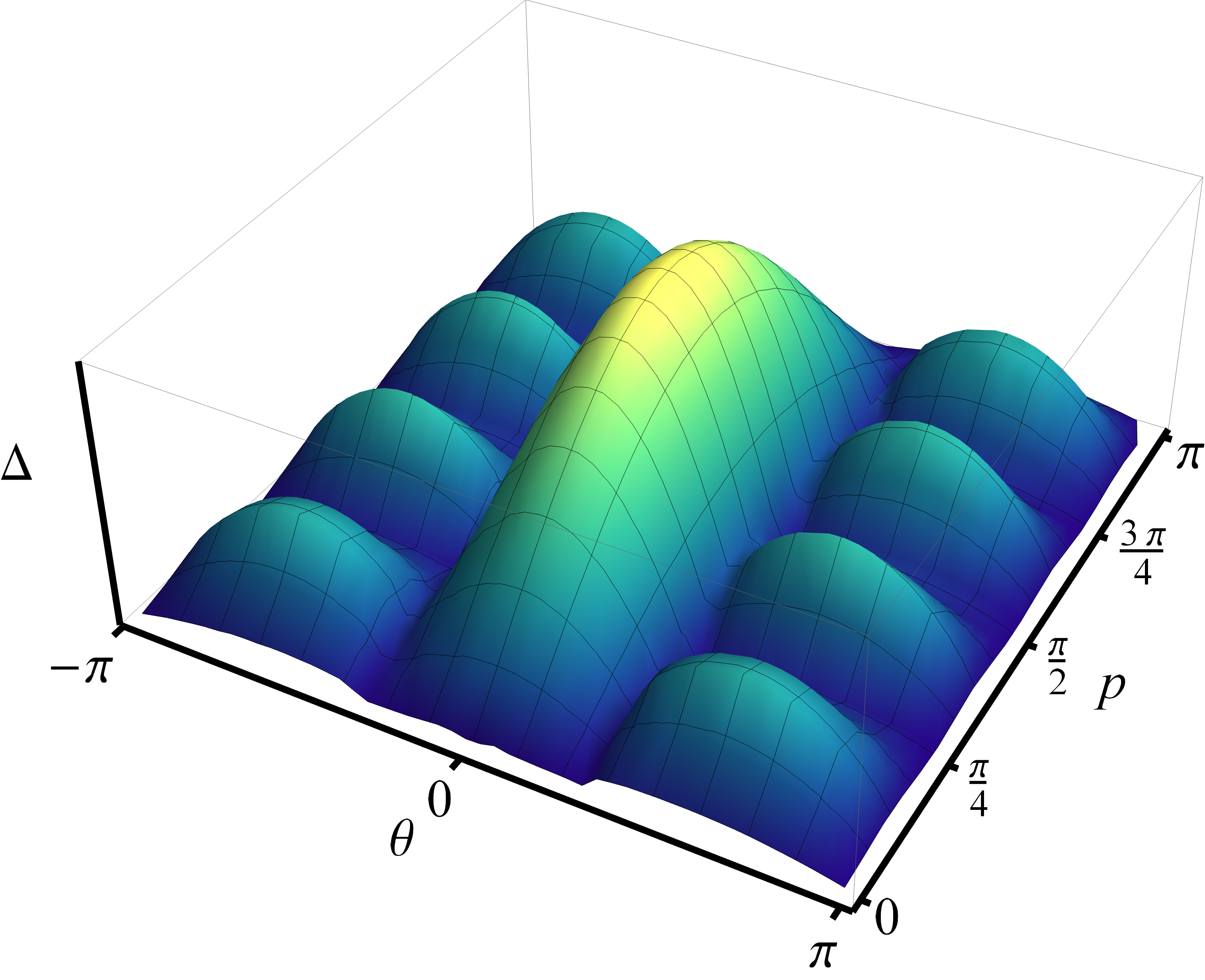}
    \caption{Energies of excited states for the chain in the thermodynamic limit:
    between momentum zero and $\pi$, two distinct regions can be observed.  The symmetry
    around zero is apparent, as well as a periodicity for the high $|\theta|$ regions,
    which indicates a breakdown of translation invariance.}
    \label{fig:excitations}
\end{figure}

Tangent-plane methods allow for the easy calculation of the energies of low-lying excited states via an MPS ansatz for
excited states
\cite{haegeman:2012a}.  The resulting lowest energies are visualized in
Fig.~\ref{fig:excitations}.

\begin{figure}[htpb]
    \centering
    \subfloat[[][]{
        \includegraphics[width=\linewidth]{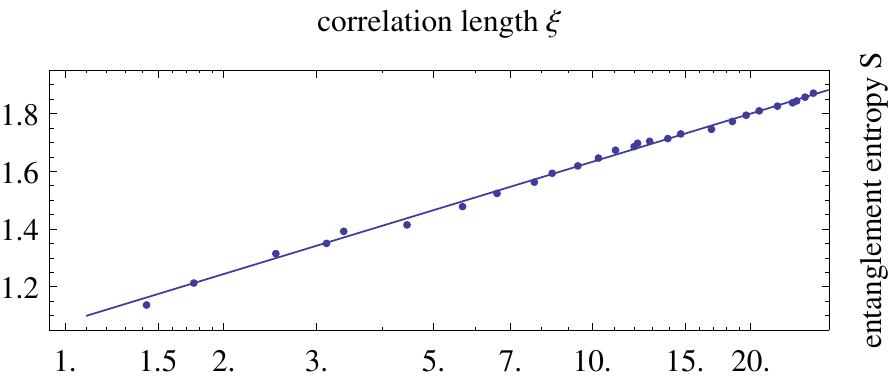}
        \label{subfig:entropycurve}
    } \\
    \subfloat[[][]{
        \includegraphics[width=\linewidth]{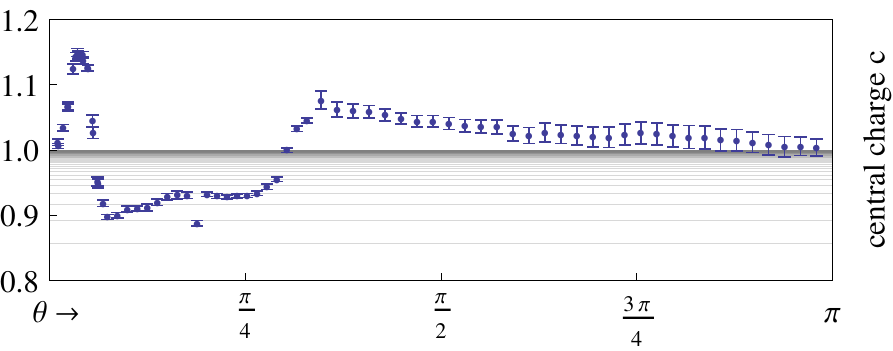}
        \label{subfig:centralcharges}
    }
    \caption[]{
        \subref{subfig:entropycurve} shows an exemplary entropy curve for
        $\theta = 3.08$ which lies within the $c=1$ region. It also includes a
        fit with $c=1.0035\pm0.012$.  \subref{subfig:centralcharges} shows
        fitted values for the central charge. The error bars included in the
        figure represent only the standard deviation based on the fit and do not
        include errors in the state, for example due to reaching a local minimum
        within the variational approach.  The horizontal lines mark allowed
        central charges $c<1$ for the unitary minimal models.
    }
    \label{fig:centralcharges}
\end{figure}

Conformal invariance allows one to characterize the collective behaviour of the
anyons for parameters where the model supports massless excitations in terms
of the central charge $c$ of the underlying Virasoro algebra.  From our
numerical data it
%
%
may be extracted from the
\emph{finite-entanglement scaling} behaviour (which fulfills a role comparable
to that of \emph{finite-size scaling}) of the entanglement entropy $S$ as a
function of the correlation length $\xi$.  Although both quantities diverge
for the exact ground state of a critical system, they remain finite for a
uniform MPS approximation at finite bond dimension, which enforces exponential
decay of correlations.  Their scaling has been shown to be determined by the
central charge \cite{fes, conformal_fes} and is described by \cite{Cardy2004}
\begin{equation}
    \label{eq:entropy}
    S(\xi) \propto \frac{c}{6} \log{\xi},
\end{equation}
where $S$ and $\xi$ are functions of $D$ that are easily obtained from the MPS
representation.  An exemplary result for this scaling behaviour is shown in
Fig.~\ref{subfig:entropycurve}.  We use this relation to estimate central
charges for the model (Fig.~\ref{subfig:centralcharges}).

\begin{figure}[htp]
    \centering
    \begin{tikzpicture}[scale=3,
            point/.style={draw,scale=0.7,thick,fill=white},
            trans/.style={draw,scale=0.7,thick,fill=white,circle},
            labell/.style={inner sep=7pt,left},
            labelr/.style={inner sep=7pt,right}
        ]

        \draw[thick] (0,0) -- (0:1) arc (0:180:1) -- cycle;
        \draw[thick] (0,0) -- (180:1) arc (180:360:1) -- cycle;
        \draw[thick,fill=blue!20] (0,0) -- (60:1) arc (60:180:1) -- cycle;
        \draw[thick,fill=blue!20] (0,0) -- (180:1) arc (180:300:1) -- cycle;
        \draw[thick,fill=red!20,dashed] (0,0) -- (5:1) arc (5:10:1) -- cycle;
        \draw[thick,fill=red!20,dashed] (0,0) -- (350:1) arc (350:355:1) -- cycle;
        \draw[thick,fill=green!20] (0,0) -- (-2:1) arc (-2:2:1) -- cycle;
        \node at (120:0.6) {$c=1$};
        \node at (35:0.6) {$\mathcal{X}$};
        \node [trans] at (0,0) {};
        \node [point] at (0:1) {};
        \node [point] at (6:1) {};
        \node [trans] at (60:1) {};
        \node [point] at (174:1) {};
        \node [trans] at (180:1) {};
        \node [point] at (186:1) {};
        \node [trans] at (300:1) {};
        \node [point] at (354:1) {};
        \node[inner sep=1.5pt] (a) at (335:0.85) {$\mathcal{Y}$};
        \node (b) at (30:1.3) {$c=\frac{8}{7}$};
        \node [labelr] at (0:1) {$\theta=0$};
        \node [labelr] at (6:1) {$\eta$};
        \node [labell] at (174:1) {$\pi\!-\!\eta$};
        \node [labell] at (180:1) {$\pi$};
        \node [labell] at (186:1) {$\pi\!+\!\eta$};
        \node [labelr] at (354:1) {$-\eta$};

        \draw (0:0.7) -- (a);
        \draw (7.5:0.8) -- (b);
        \draw (352.5:0.8) -- (b);

    \end{tikzpicture}
    \caption{Phase diagram for the $SO(5)_2$ $\psi_3$-chain derived from
        numerical (MPS) results and including integrable points at $|\theta| =
        0,\eta,\pi\!-\!\eta, \pi$.  We do not have consistent results for region
        $\mathcal{X}$, where the central charge appears to be less than one.
        Bethe ansatz results indicate that region $\mathcal{Y}$ is gapped.
    }
    \label{fig:phasediagram}
\end{figure}
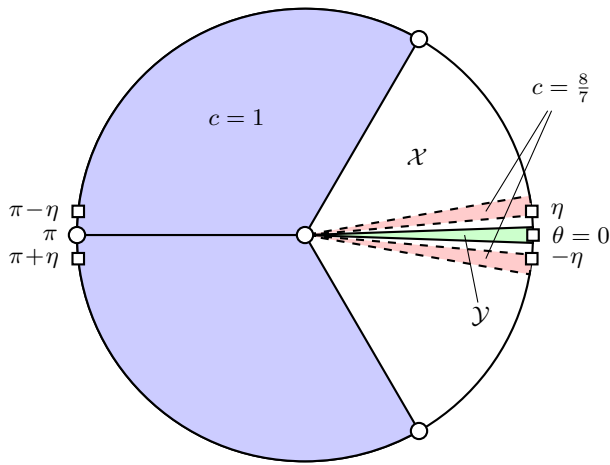

\noindent
Using the central charge and dispersion relations we obtained a phase portrait
(Fig.~\ref{fig:phasediagram}), from which we see that there exist two large
regions, $\frac{\pi}{3}\lesssim |\theta|<\pi$,
with central charge $c=1$ and two small intervals near
$\theta\approx\pm0.033\pi$ where the central charge of the model is found to
be $c=8/7$.  These results agree with the analytic results at two of the
integrable points discussed below.
Beyond these regions we got spurious results, possibly due to local energy
minima, or due to especially strong corrections to the asymptotic relation
\eqref{eq:entropy}, and it was not possible to determine a consistent central
charge.
Between the $c=1$ sectors at $\theta=\pi$ the spectrum shows level crossings
with large degeneracies, indicating a first order transition. This is consistent
with a fixed point of the $\psi_2 \leftrightarrow \psi_5$ automorphism.
Finally, near
$\theta=0$, our analytical approach predicts a small gapped region (see
below).

\paragraph{Integrable points.}
The numerical investigation can be complemented by analytical results for
special values of the coupling parameter for which the model becomes
integrable.  The identification of integrable points and associated
$R$-matrices solving the Yang-Baxter equation (YBE) is achieved though the
realization that the projection operators (\ref{eqnproj}) form a
representation of the Birman-Murakami-Wenzl (BMW) algebra
\cite{{BiWe89,*Mura87}}.
%
In particular, choosing $\theta=0,\pi$, the Hamiltonian can be written as
\begin{equation*}
  \mathcal{H}_{0,\pi}  =  
    \frac{\mp1}{\sqrt{10}} \sum_{i=1}^\L u_{i}+\mathrm{const.}\,, 
\end{equation*}
where the $u_{i} = \sqrt{5}\,p_{i}^{(1)}$ generate a representation of a
subalgebra of the BMW algebra isomorphic to the periodic Temperley-Lieb
algebra \cite{TeLi71}.  This model can be derived from an $R$-matrix
\begin{equation}
\label{rmat}
  R(u)  =  w_{1}(u)p^{(1)} + w_{2}(u)p^{(2)} + w_{5}(u)p^{(5)},
\end{equation}
with weights $w_{1}(u) = \sinh(\gamma-u)$ and $w_{2}(u) =\sinh(\gamma+u)=
w_{5}(u)$, where $\cosh\gamma = \frac{\sqrt{5}}{2}$.  The global Hamiltonian
(\ref{eqhamil}) can be obtained as the logarithmic derivative of a commuting
transfer matrix built from this $R$-matrix.

As a consequence of the underlying Temperley-Lieb algebra the spectrum of the
model at $\theta=0,\pi$ can be related to that of the XXZ spin-1/2 chain with
anisotropy $\pm\Delta=\cosh\gamma=\frac{\sqrt{5}}{2}$ \cite{TeLi71,OwBa87}.
This implies that the eigenstates of the model can be parametrized by
solutions $\{u_{j}\}_{j=1}^{n}$ to the Bethe equations associated of the XXZ
model for suitably twisted boundary conditions
\begin{equation*}
\left[\frac{\sinh(u_{j}-\frac{\gamma}{2})}{
    \sinh(u_{j}+\frac{\gamma}{2})}\right]^{\L} 
=  - \zeta^{-2} \prod_{k=1}^{n}
\frac{\sinh(u_{j}-u_{k}-\gamma)}{\sinh(u_{j}-u_{k}+\gamma)}. 
\end{equation*}
The twist $\zeta$ appearing in the Bethe equations depends on the sector $n$:
for $n=\frac{\L}{2}$ it can take values $\pm
i,e^{\frac{\pm2i\pi}{3}},e^{\pm\gamma}$ while $\zeta^{\L-\frac{n}{2}}=1$
otherwise.  This is similar to other higher spin Temperley-Lieb chains
\cite{AuKl10}.
The bulk properties of the quantum chain can be obtained from the equivalence
to the XXZ chain: 
$\theta=0$ corresponds to the anti-ferromagnetic spin chain which has a tiny
energy gap of $\Delta = {\pi} \exp({-\frac{\pi^{2}}{2\gamma}}) /({\gamma
  \sqrt{10}})\simeq 2.9\,\times 10^{-4}$ \cite{ClGa66}.  This gap is too small to be
resolved in our numerical results but based on continuity arguments we expect
that there is an extended massive region in the phase diagram of the anyon
chain around $\theta=0$.
As a consequence of the multiple twists in the anyon chain the
thermodynamic ground state of this model is tenfold degenerate.
%

At $\theta=\pi$ the Temperley-Lieb equivalence is to the massive ferromagnetic
XXZ chain.  The ground states have energy $E = -\frac{1}{\sqrt{2}} \L$ and all
possible momenta.  We also find that there is a large range of excitations
lying above the gap which are characterized by the parameters $l,k \in
\Z$. The energy and momentum of these states are given by
\begin{equation*}
\begin{aligned}
  \Delta E  &=  \frac{1}{\sqrt{2}}- \frac{2}{\sqrt{10}}
  \cos\left(\frac{2k\pi}{\L}-\frac{4l\pi}{\L(\L-2)}\right), \\ 
  P  &=  \frac{2\pi}{\L}\left(k + l\right).
\end{aligned}
\end{equation*}
As observed in the numerical analysis the spectrum of the anyon chain shows
degeneracies growing with the system size indicating a first order transition
at $\theta=\pi$.

We have identified two additional pairs of coupling constants where the anyon
model becomes integrable: for $\theta=\eta, \pi+\eta$ with $\eta =
-\frac{\pi}{4} + \mbox{atan}\left(\frac{1+\sqrt{5}}{4}\right)$ the Hamiltonian
(\ref{eqhamil}) can be obtained from an $R$-matrix associated with the
\emph{full} BMW algebra which is given by (\ref{rmat}) with weights
\begin{equation}
\label{rweight2}
\begin{aligned}
  w_1(u) &= \sinh\left(u+\frac{i\pi}{10}\right)
    \sinh\left(u+\frac{3i\pi}{10}\right) = w_2(-u)\,,\\
  w_5(u) &= \sinh\left(u+\frac{9i\pi}{10}\right)
    \sinh\left(u+\frac{3i\pi}{10}\right)\,.
\end{aligned}
\end{equation}
A third solution to the YBE is related to (\ref{rweight2}) by the
transformation mapping $\theta \leftrightarrow -\theta$ mentioned above. 

For both of these solutions we can construct a set of six commuting transfer
matrices $t^{(\ell)}(u)$, $\ell=1,\ldots,6$, whose asymptotics as $u\to\pm\infty$ is
given by the topological charges (\ref{topoy}) up to a scalar function.
%
This implies that the charges $Y_\ell$ are elements of the algebra of
commuting integrals of the anyon chain.  By diagonalizing the transfer matrix
$t^{(3)}(u)$ we find that the spectrum of the anyon chain at these integrable
points can be obtained from
\begin{equation}
\label{bethe2}
  \left[i\frac{\sinh\left(u_{j}+\frac{i\pi}{20}\right)}{
      \sinh\left(u_{j}-\frac{i\pi}{20}\right)} \right]^{\L}
  = - \zeta \prod_{k=1}^{n}
    \frac{\sinh(u_{j}-u_{k}+\frac{2i\pi}{5})}{
      \sinh(u_{j}-u_{k}-\frac{2i\pi}{5})}\,
\end{equation}
where the $\zeta=\pm1$ is the eigenvalue of $Y_6$.  Note that these Bethe
equations are (up to a twist) those of the $\Z_{5}$ Fateev-Zamolodchikov (FZ)
model, sometimes also referred to as the five-state self-dual chiral Potts model
\cite{FaZa82,Albe94}.  This can be understood as a consequence that the
$R$-matrices with (\ref{rweight2}) correspond to descendants of the zero-field
six-vertex model \cite{BaSt90}.

Solving the Bethe equations (\ref{bethe2}) we have analyzed the ground state
and low lying excitations of the anyon chain in the thermodynamic limit and
found that these integrable points are described by effective conformal field
theories (CFTs) with central charges $c=\frac{8}{7}$ for $\theta=\pm\eta$ and
$c=1$ for $\pi\pm\eta$, just as the ferro- (antiferro-)magnetic $\Z_{5}$ FZ
model \cite{Albe94}.  
Based on the $S$-matrix (\ref{smat}) and the operator content of the anyon
chain as identified from our finite-size analysis of the spectrum we
conjecture, however, that the low energy theories are unitary rational models
invariant under extensions of the Virasoro algebra, i.e.\ the $\mathcal{W}B_2$
($\mathcal{W}D_5$) algebra for $\theta=\pm\eta$ ($\pi\pm\eta$).
An extended and more formal treatment of the integrable points will be
presented in a separate publication \cite{FiFF14}.

\paragraph{Conclusion.} 
We introduced a lattice model of the anyons expected to be relevant for the
low-energy physics of the $M=5$ topological Kondo effect.  The degeneracy of
the anyon zero modes is lifted by local interactions consistent by the
$SO(5)_2$ fusion rules.
We have been able to characterize the collective states of the resulting
strongly interacting anyon model
by a combination of numerical simulation and
analytic results.  We found multiple extended critical regions with central
charges $c=1$ and $c=8/7$, which were calculated from finite-entanglement
scaling.  In addition we were able to identify the low-energy effective field
theory for special values of the coupling from the Bethe ansatz solution.

\begin{acknowledgments}
  The authors would like to thank Eddy Ardonne and Michael Flohr for helpful
  discussions.  This work was supported by the Deutsche Forschungsgemeinschaft
  under grant no.  Fr-737/7, by the ERC grants QFTCMPS and SIQS, and by the
  cluster of excellence EXC 201 Quantum Engineering and Space-Time Research.
\end{acknowledgments}



\begin{thebibliography}{8}%
\makeatletter
\providecommand \@ifxundefined [1]{%
 \@ifx{#1\undefined}
}%
\providecommand \@ifnum [1]{%
 \ifnum #1\expandafter \@firstoftwo
 \else \expandafter \@secondoftwo
 \fi
}%
\providecommand \@ifx [1]{%
 \ifx #1\expandafter \@firstoftwo
 \else \expandafter \@secondoftwo
 \fi
}%
\providecommand \natexlab [1]{#1}%
\providecommand \enquote  [1]{``#1''}%
\providecommand \bibnamefont  [1]{#1}%
\providecommand \bibfnamefont [1]{#1}%
\providecommand \citenamefont [1]{#1}%
\providecommand \href@noop [0]{\@secondoftwo}%
\providecommand \href [0]{\begingroup \@sanitize@url \@href}%
\providecommand \@href[1]{\@@startlink{#1}\@@href}%
\providecommand \@@href[1]{\endgroup#1\@@endlink}%
\providecommand \@sanitize@url [0]{\catcode `\\12\catcode `\$12\catcode
  `\&12\catcode `\#12\catcode `\^12\catcode `\_12\catcode `\%12\relax}%
\providecommand \@@startlink[1]{}%
\providecommand \@@endlink[0]{}%
\providecommand \url  [0]{\begingroup\@sanitize@url \@url }%
\providecommand \@url [1]{\endgroup\@href {#1}{\urlprefix }}%
\providecommand \urlprefix  [0]{URL }%
\providecommand \Eprint [0]{\href }%
\providecommand \doibase [0]{http://dx.doi.org/}%
\providecommand \selectlanguage [0]{\@gobble}%
\providecommand \bibinfo  [0]{\@secondoftwo}%
\providecommand \bibfield  [0]{\@secondoftwo}%
\providecommand \translation [1]{[#1]}%
\providecommand \BibitemOpen [0]{}%
\providecommand \bibitemStop [0]{}%
\providecommand \bibitemNoStop [0]{.\EOS\space}%
\providecommand \EOS [0]{\spacefactor3000\relax}%
\providecommand \BibitemShut  [1]{\csname bibitem#1\endcsname}%
\let\auto@bib@innerbib\@empty
\bibitem [{\citenamefont {Wilczek}(1990)}]{wilczek:1990a}%
  \BibitemOpen
  \bibfield  {author} {\bibinfo {author} {\bibfnamefont {F.}~\bibnamefont
  {Wilczek}},\ }\href@noop {} {\emph {\bibinfo {title} {Fractional statistics
  and anyon superconductivity}}}\ (\bibinfo  {publisher} {World Scientific
  Publishing Co. Inc.},\ \bibinfo {address} {Teaneck, NJ},\ \bibinfo {year}
  {1990})\ pp.\ \bibinfo {pages} {x+447}\BibitemShut {NoStop}%
\bibitem [{\citenamefont {Moore}\ and\ \citenamefont {Read}(1991)}]{Moore1991}%
  \BibitemOpen
  \bibfield  {author} {\bibinfo {author} {\bibfnamefont {G.}~\bibnamefont
  {Moore}}\ and\ \bibinfo {author} {\bibfnamefont {N.}~\bibnamefont {Read}},\
  }\href {\doibase 10.1016/0550-3213(91)90407-O} {\bibfield  {journal}
  {\bibinfo  {journal} {Nucl. Phys. B}\ }\textbf {\bibinfo {volume}
  {360}},\ \bibinfo {pages} {362} (\bibinfo {year} {1991})}\BibitemShut
  {NoStop}%
\bibitem [{\citenamefont {Read}\ and\ \citenamefont {Rezayi}(1999)}]{Read1999}%
  \BibitemOpen
  \bibfield  {author} {\bibinfo {author} {\bibfnamefont {N.}~\bibnamefont
  {Read}}\ and\ \bibinfo {author} {\bibfnamefont {E.}~\bibnamefont {Rezayi}},\
  }\href {\doibase 10.1103/PhysRevB.59.8084} {\bibfield  {journal} {\bibinfo
  {journal} {Phys. Rev. B}\ }\textbf {\bibinfo {volume} {59}},\ \bibinfo
  {pages} {8084} (\bibinfo {year} {1999})}\BibitemShut {NoStop}%
\bibitem [{\citenamefont {Stanescu}\ \emph {et~al.}(2011)\citenamefont
  {Stanescu}, \citenamefont {Lutchyn},\ and\ \citenamefont
  {Das~Sarma}}]{Stanescu2011}%
  \BibitemOpen
  \bibfield  {author} {\bibinfo {author} {\bibfnamefont {T.~D.}\ \bibnamefont
  {Stanescu}}, \bibinfo {author} {\bibfnamefont {R.~M.}\ \bibnamefont
  {Lutchyn}}, \ and\ \bibinfo {author} {\bibfnamefont {S.}~\bibnamefont
  {Das~Sarma}},\ }\href {\doibase 10.1103/PhysRevB.84.144522} {\bibfield
  {journal} {\bibinfo  {journal} {Phys. Rev. B}\ }\textbf {\bibinfo
  {volume} {84}} (\bibinfo {year} {2011})}\BibitemShut {NoStop}%
\bibitem [{\citenamefont {Mourik}\ \emph {et~al.}(2012)\citenamefont {Mourik},
  \citenamefont {Zuo}, \citenamefont {Frolov}, \citenamefont {Plissard},
  \citenamefont {Bakkers},\ and\ \citenamefont {Kouwenhoven}}]{MZFP12}%
  \BibitemOpen
  \bibfield  {author} {\bibinfo {author} {\bibfnamefont {V.}~\bibnamefont
  {Mourik}}, \bibinfo {author} {\bibfnamefont {K.}~\bibnamefont {Zuo}},
  \bibinfo {author} {\bibfnamefont {S.~M.}\ \bibnamefont {Frolov}}, \bibinfo
  {author} {\bibfnamefont {S.~R.}\ \bibnamefont {Plissard}}, \bibinfo {author}
  {\bibfnamefont {E.~P. A.~M.}\ \bibnamefont {Bakkers}}, \ and\ \bibinfo
  {author} {\bibfnamefont {L.~P.}\ \bibnamefont {Kouwenhoven}},\ }\href@noop {}
  {\bibfield  {journal} {\bibinfo  {journal} {Science}\ }\textbf {\bibinfo
  {volume} {336}},\ \bibinfo {pages} {1003} (\bibinfo {year} {2012})},\ \Eprint
  {http://arxiv.org/abs/1204.2792} {arXiv:1204.2792} \BibitemShut {NoStop}%
\bibitem [{\citenamefont {Read}\ and\ \citenamefont {Green}(2000)}]{Read2000}%
  \BibitemOpen
  \bibfield  {author} {\bibinfo {author} {\bibfnamefont {N.}~\bibnamefont
  {Read}}\ and\ \bibinfo {author} {\bibfnamefont {D.}~\bibnamefont {Green}},\
  }\href {\doibase 10.1103/PhysRevB.61.10267} {\bibfield  {journal} {\bibinfo
  {journal} {Phys. Rev. B}\ }\textbf {\bibinfo {volume} {61}},\ \bibinfo
  {pages} {10267} (\bibinfo {year} {2000})}\BibitemShut {NoStop}%
\bibitem [{\citenamefont {Dennis}\ \emph {et~al.}(2002)\citenamefont {Dennis},
  \citenamefont {Kitaev}, \citenamefont {Landahl},\ and\ \citenamefont
  {Preskill}}]{dennis:2002a}%
  \BibitemOpen
  \bibfield  {author} {\bibinfo {author} {\bibfnamefont {E.}~\bibnamefont
  {Dennis}}, \bibinfo {author} {\bibfnamefont {A.}~\bibnamefont {Kitaev}},
  \bibinfo {author} {\bibfnamefont {A.}~\bibnamefont {Landahl}}, \ and\
  \bibinfo {author} {\bibfnamefont {J.}~\bibnamefont {Preskill}},\ }\href
  {\doibase dx.doi.org/10.1063/1.1499754} 
  {\bibfield  {journal} {\bibinfo  {journal} {J. Math. Phys.}\ }\textbf
  {\bibinfo {volume} {43}},\ \bibinfo {pages} {4452} (\bibinfo {year}
  {2002})}\BibitemShut {NoStop}%
\bibitem [{\citenamefont {Nayak}\ \emph {et~al.}(2008)\citenamefont {Nayak},
  \citenamefont {Simon}, \citenamefont {Stern}, \citenamefont {Freedman},\ and\
  \citenamefont {Das~Sarma}}]{nayak:2008a}%
  \BibitemOpen
  \bibfield  {author} {\bibinfo {author} {\bibfnamefont {C.}~\bibnamefont
  {Nayak}}, \bibinfo {author} {\bibfnamefont {S.~H.}\ \bibnamefont {Simon}},
  \bibinfo {author} {\bibfnamefont {A.}~\bibnamefont {Stern}}, \bibinfo
  {author} {\bibfnamefont {M.}~\bibnamefont {Freedman}}, \ and\ \bibinfo
  {author} {\bibfnamefont {S.}~\bibnamefont {Das~Sarma}},\ }\href {\doibase
  10.1103/RevModPhys.80.1083} {\bibfield  {journal} {\bibinfo  {journal} {Rev.
  Mod. Phys.}\ }\textbf {\bibinfo {volume} {80}},\ \bibinfo {pages} {1083 }
  (\bibinfo {year} {2008})}\BibitemShut {NoStop}%
\bibitem [{\citenamefont {Kitaev}(2003)}]{kitaev:2003a}%
  \BibitemOpen
  \bibfield  {author} {\bibinfo {author} {\bibfnamefont {A.~Y.}\ \bibnamefont
  {Kitaev}},\ }\href {\doibase 10.1016/S0003-4916(02)00018-0} 
  {\bibfield  {journal} {\bibinfo  {journal} {Ann.
  Phys.}\ }\textbf {\bibinfo {volume} {303}},\ \bibinfo {pages} {2 } (\bibinfo
  {year} {2003})}\BibitemShut {NoStop}%
\bibitem [{\citenamefont {Kitaev}(2006)}]{kitaev:2006a}%
  \BibitemOpen
  \bibfield  {author} {\bibinfo {author} {\bibfnamefont {A.}~\bibnamefont
  {Kitaev}},\ }\href {\doibase 10.1016/j.aop.2005.10.005} 
  {\bibfield  {journal} {\bibinfo  {journal} {Ann.
  Phys.}\ }\textbf {\bibinfo {volume} {321}},\ \bibinfo {pages} {2 } (\bibinfo
  {year} {2006})}\ 
  \BibitemShut {NoStop}%
\bibitem [{\citenamefont {Wang}(2008)}]{wang:2008a}%
  \BibitemOpen
  \bibfield  {author} {\bibinfo {author} {\bibfnamefont {Z.}~\bibnamefont
  {Wang}},\ }\href@noop {} {\emph {\bibinfo {title} {Topological {Q}uantum
  {C}omputation}}}\ (\bibinfo  {publisher} {American Mathematical Society},\
  \bibinfo {year} {2008})\BibitemShut {NoStop}%
\bibitem [{pac(2012)}]{pachos:2012a}%
  \BibitemOpen
  \href@noop {} {\emph {\bibinfo {title} {Introduction to {T}opological
  {Q}uantum {C}omputation}}}\ (\bibinfo  {publisher} {Cambridge University
  Press},\ \bibinfo {address} {Cambridge},\ \bibinfo {year} {2012})\BibitemShut
  {NoStop}%
\bibitem [{\citenamefont {Barends}\ \emph {et~al.}(2014)\citenamefont
  {Barends}, \citenamefont {Kelly}, \citenamefont {Megrant}, \citenamefont
  {Veitia}, \citenamefont {Sank}, \citenamefont {Jeffrey}, \citenamefont
  {White}, \citenamefont {Mutus}, \citenamefont {Fowler}, \citenamefont
  {Campbell}, \citenamefont {Chen}, \citenamefont {Chen}, \citenamefont
  {Chiaro}, \citenamefont {Dunsworth}, \citenamefont {Neill}, \citenamefont
  {O'Malley}, \citenamefont {Roushan}, \citenamefont {Vainsencher},
  \citenamefont {Wenner}, \citenamefont {Korotkov}, \citenamefont {Cleland},\
  and\ \citenamefont {Martinis}}]{barends:2014a}%
  \BibitemOpen
  \bibfield  {author} {\bibinfo {author} {\bibfnamefont {R.}~\bibnamefont
  {Barends}}\ \emph {et~al.},\ }
  \Eprint {http://arxiv.org/abs/1402.4848} {arXiv:1402.4848}\
  (\bibinfo {year} {2014})\BibitemShut {NoStop}%
\bibitem{beenakker:2013a} C. W. J. Beenakker, Annu. Rev. Con. Mat. Phys. {\bf 4}, 113 (2013).
\bibitem [{\citenamefont {B{\'e}ri}\ and\ \citenamefont
  {Cooper}(2012)}]{BeCo12}%
  \BibitemOpen
  \bibfield  {author} {\bibinfo {author} {\bibfnamefont {B.}~\bibnamefont
  {B{\'e}ri}}\ and\ \bibinfo {author} {\bibfnamefont {N.~R.}\ \bibnamefont
  {Cooper}},\ }\href@noop {} {\bibfield  {journal} {\bibinfo  {journal} {Phys.
  Rev. Lett.}\ }\textbf {\bibinfo {volume} {109}},\ \bibinfo {pages} {156803}
  (\bibinfo {year} {2012})},\ \Eprint {http://arxiv.org/abs/1206.2224}
  {arXiv:1206.2224} \BibitemShut {NoStop}%
\bibitem [{\citenamefont {Altland}\ \emph {et~al.}(2013)\citenamefont
  {Altland}, \citenamefont {B{\'e}ri}, \citenamefont {Egger},\ and\
  \citenamefont {Tsvelik}}]{ABET13}%
  \BibitemOpen
  \bibfield  {author} {\bibinfo {author} {\bibfnamefont {A.}~\bibnamefont
  {Altland}}, \bibinfo {author} {\bibfnamefont {B.}~\bibnamefont {B{\'e}ri}},
  \bibinfo {author} {\bibfnamefont {R.}~\bibnamefont {Egger}}, and \bibinfo
  {author} {\bibfnamefont {A.~M.}\ \bibnamefont {Tsvelik}},\ }\href@noop {}
  {\bibfield  {journal} {\bibinfo  {journal} {preprint}\ } (\bibinfo {year}
  {2013})},\ \Eprint {http://arxiv.org/abs/1312.3802} {1312.3802} \BibitemShut
  {NoStop}%
\bibitem [{\citenamefont {Altland}\ \emph {et~al.}(2014)\citenamefont
  {Altland}, \citenamefont {B{\'e}ri}, \citenamefont {Egger}, and
  \citenamefont {Tsvelik}}]{ABET14}%
  \BibitemOpen
  \bibfield  {author} {\bibinfo {author} {\bibfnamefont {A.}~\bibnamefont
  {Altland}}, \bibinfo {author} {\bibfnamefont {B.}~\bibnamefont {B{\'e}ri}},
  \bibinfo {author} {\bibfnamefont {R.}~\bibnamefont {Egger}}, and \bibinfo
  {author} {\bibfnamefont {A.~M.}\ \bibnamefont {Tsvelik}},\ }\href@noop {}
  {\bibfield  {journal} {\bibinfo  {journal} {J. Phys. A}\ }\textbf
  {\bibinfo {volume} {47}},\ \bibinfo {pages} {265001} (\bibinfo {year}
  {2014})} \BibitemShut
  {NoStop}%
\bibitem [{\citenamefont {Cui}\ and\ \citenamefont {Wang}(2014)}]{cui:2014a}%
  \BibitemOpen
  \bibfield  {author} {\bibinfo {author} {\bibfnamefont {S.~X.}\ \bibnamefont
  {Cui}}\ and\ \bibinfo {author} {\bibfnamefont {Z.}~\bibnamefont {Wang}},\
  } \Eprint {http://arxiv.org/abs/1405.7778} {arXiv:1405.7778}\
  (\bibinfo {year} {2014})\BibitemShut {NoStop}%
\bibitem [{\citenamefont {Andrews}\ \emph {et~al.}(1984)\citenamefont
  {Andrews}, \citenamefont {Baxter},\ and\ \citenamefont {Forrester}}]{AnBF84}%
  \BibitemOpen
  \bibfield  {author} {\bibinfo {author} {\bibfnamefont {G.~E.}\ \bibnamefont
  {Andrews}}, \bibinfo {author} {\bibfnamefont {R.~J.}\ \bibnamefont {Baxter}},
  \ and\ \bibinfo {author} {\bibfnamefont {P.~J.}\ \bibnamefont {Forrester}},\
  }\href {\doibase 10.1007/BF01014383} 
  {\bibfield  {journal} {\bibinfo  {journal} {J. Stat. Phys.}\
  }\textbf {\bibinfo {volume} {35}},\ \bibinfo {pages} {193} (\bibinfo {year}
  {1984})}\BibitemShut {NoStop}%
\bibitem [{\citenamefont {Kakashvili}\ and\ \citenamefont
  {Ardonne}(2012)}]{Kakashvili2012}%
  \BibitemOpen
  \bibfield  {author} {\bibinfo {author} {\bibfnamefont {P.}~\bibnamefont
  {Kakashvili}}\ and\ \bibinfo {author} {\bibfnamefont {E.}~\bibnamefont
  {Ardonne}},\ }\href {\doibase 10.1103/PhysRevB.85.115116} {\bibfield
  {journal} {\bibinfo  {journal} {Phys. Rev. B}\ }\textbf {\bibinfo
  {volume} {85}}, \bibinfo {pages} {115116} (\bibinfo {year} {2012})}\BibitemShut {NoStop}%
\bibitem [{\citenamefont {Feiguin}\ \emph {et~al.}(2007)\citenamefont
  {Feiguin}, \citenamefont {Trebst}, \citenamefont {Ludwig}, \citenamefont
  {Troyer}, \citenamefont {Kitaev}, \citenamefont {Wang},\ and\ \citenamefont
  {Freedman}}]{feiguin:2007a}%
  \BibitemOpen
  \bibfield  {author} {\bibinfo {author} {\bibfnamefont {A.}~\bibnamefont
  {Feiguin}}, \bibinfo {author} {\bibfnamefont {S.}~\bibnamefont {Trebst}},
  \bibinfo {author} {\bibfnamefont {A.~W.~W.}\ \bibnamefont {Ludwig}}, \bibinfo
  {author} {\bibfnamefont {M.}~\bibnamefont {Troyer}}, \bibinfo {author}
  {\bibfnamefont {A.}~\bibnamefont {Kitaev}}, \bibinfo {author} {\bibfnamefont
  {Z.}~\bibnamefont {Wang}}, \ and\ \bibinfo {author} {\bibfnamefont {M.~H.}\
  \bibnamefont {Freedman}},\ }\href {\doibase 10.1103/PhysRevLett.98.160409} 
  {\bibfield  {journal} {\bibinfo
  {journal} {Phys. Rev. Lett.}\ }\textbf {\bibinfo {volume} {98}},\ \bibinfo
  {pages} {160409} (\bibinfo {year} {2007})}\BibitemShut {NoStop}%
\bibitem [{\citenamefont {Gils}\ \emph {et~al.}(2013)\citenamefont {Gils},
  \citenamefont {Ardonne}, \citenamefont {Trebst}, \citenamefont {Huse},
  \citenamefont {Ludwig}, \citenamefont {Troyer},\ and\ \citenamefont
  {Wang}}]{GATH13}%
  \BibitemOpen
  \bibfield  {author} {\bibinfo {author} {\bibfnamefont {C.}~\bibnamefont
  {Gils}}, \bibinfo {author} {\bibfnamefont {E.}~\bibnamefont {Ardonne}},
  \bibinfo {author} {\bibfnamefont {S.}~\bibnamefont {Trebst}}, \bibinfo
  {author} {\bibfnamefont {D.~A.}\ \bibnamefont {Huse}}, \bibinfo {author}
  {\bibfnamefont {A.~W.}\ \bibnamefont {Ludwig}}, \bibinfo {author}
  {\bibfnamefont {M.}~\bibnamefont {Troyer}}, \ and\ \bibinfo {author}
  {\bibfnamefont {Z.}~\bibnamefont {Wang}},\ }\href
  {http://prb.aps.org/abstract/PRB/v87/i23/e235120} {\bibfield  {journal}
  {\bibinfo  {journal} {Physical Review B}\ }\textbf {\bibinfo {volume} {87}},\
  \bibinfo {pages} {235120} (\bibinfo {year} {2013})}\BibitemShut {NoStop}%
\bibitem [{\citenamefont {Gils}\ \emph {et~al.}(2009)\citenamefont {Gils},
  \citenamefont {Ardonne}, \citenamefont {Trebst}, \citenamefont {Ludwig},
  \citenamefont {Troyer},\ and\ \citenamefont {Wang}}]{Gils2009}%
  \BibitemOpen
  \bibfield  {author} {\bibinfo {author} {\bibfnamefont {C.}~\bibnamefont
  {Gils}}, \bibinfo {author} {\bibfnamefont {E.}~\bibnamefont {Ardonne}},
  \bibinfo {author} {\bibfnamefont {S.}~\bibnamefont {Trebst}}, \bibinfo
  {author} {\bibfnamefont {A.}~\bibnamefont {Ludwig}}, \bibinfo {author}
  {\bibfnamefont {M.}~\bibnamefont {Troyer}}, \ and\ \bibinfo {author}
  {\bibfnamefont {Z.}~\bibnamefont {Wang}},\ }\href {\doibase
  10.1103/PhysRevLett.103.070401} {\bibfield  {journal} {\bibinfo  {journal}
  {Physical Review Letters}\ }\textbf {\bibinfo {volume} {103}},\
  \bibinfo {pages} {070401} (\bibinfo
  {year} {2009})}\BibitemShut {NoStop}%
\bibitem [{\citenamefont {Schollw{\"o}ck}(2005)}]{schollwoeck:2005a}%
  \BibitemOpen
  \bibfield  {author} {\bibinfo {author} {\bibfnamefont {U.}~\bibnamefont
  {Schollw{\"o}ck}},\ }\href {\doibase 10.1103/RevModPhys.77.259} 
  {\bibfield  {journal} {\bibinfo  {journal}
  {Rev. Mod. Phys.}\ }\textbf {\bibinfo {volume} {77}},\ \bibinfo {pages}
  {259} (\bibinfo {year} {2005})}\ 
  \BibitemShut {NoStop}%
\bibitem [{\citenamefont {Schollw{\"o}ck}(2011)}]{schollwock:2011a}%
  \BibitemOpen
  \bibfield  {author} {\bibinfo {author} {\bibfnamefont {U.}~\bibnamefont
  {Schollw{\"o}ck}},\ }\href {\doibase 10.1016/j.aop.2010.09.012} 
  {\bibfield  {journal} {\bibinfo  {journal}
  {Ann. Phys.}\ }\textbf {\bibinfo {volume} {326}},\ \bibinfo {pages} {96 }
  (\bibinfo {year} {2011})}\BibitemShut {NoStop}%
\bibitem [{\citenamefont {Vidal}(2003)}]{vidal:2003a}%
  \BibitemOpen
  \bibfield  {author} {\bibinfo {author} {\bibfnamefont {G.}~\bibnamefont
  {Vidal}},\ }\href {\doibase 10.1103/PhysRevLett.93.040502} 
  {\bibfield  {journal} {\bibinfo  {journal} {Phys.
  Rev. Lett.}\ }\textbf {\bibinfo {volume} {93}},\ \bibinfo {pages} {040502}
  (\bibinfo {year} {2003})}\ 
  \BibitemShut
  {NoStop}%
\bibitem [{\citenamefont {Haegeman}\ \emph {et~al.}(2011)\citenamefont
  {Haegeman}, \citenamefont {Cirac}, \citenamefont {Osborne}, \citenamefont
  {Pi\ifmmode~\check{z}\else \v{z}\fi{}orn}, \citenamefont {Verschelde},\ and\
  \citenamefont {Verstraete}}]{haegeman:2011b}%
  \BibitemOpen
  \bibfield  {author} {\bibinfo {author} {\bibfnamefont {J.}~\bibnamefont
  {Haegeman}}, \bibinfo {author} {\bibfnamefont {J.~I.}\ \bibnamefont {Cirac}},
  \bibinfo {author} {\bibfnamefont {T.~J.}\ \bibnamefont {Osborne}}, \bibinfo
  {author} {\bibfnamefont {I.}~\bibnamefont {Pi\ifmmode~\check{z}\else
  \v{z}\fi{}orn}}, \bibinfo {author} {\bibfnamefont {H.}~\bibnamefont
  {Verschelde}}, \ and\ \bibinfo {author} {\bibfnamefont {F.}~\bibnamefont
  {Verstraete}},\ }\href {\doibase 10.1103/PhysRevLett.107.070601} {\bibfield
  {journal} {\bibinfo  {journal} {Phys. Rev. Lett.}\ }\textbf {\bibinfo
  {volume} {107}},\ \bibinfo {pages} {070601} (\bibinfo {year} {2011})}\
  \BibitemShut {NoStop}%
\bibitem [{\citenamefont {Corboz}\ and\ \citenamefont
  {Vidal}(2009)}]{corboz:2009a}%
  \BibitemOpen
  \bibfield  {author} {\bibinfo {author} {\bibfnamefont {P.}~\bibnamefont
  {Corboz}}\ and\ \bibinfo {author} {\bibfnamefont {G.}~\bibnamefont {Vidal}},\
  }\href {\doibase 10.1103/PhysRevB.80.165129} {\bibfield  {journal} {\bibinfo
  {journal} {Phys. Rev. B}\ }\textbf {\bibinfo {volume} {80}},\ \bibinfo
  {pages} {165129} (\bibinfo {year} {2009})}\ 
  \BibitemShut {NoStop}%
\bibitem [{\citenamefont {Corboz}\ \emph
  {et~al.}(2010{\natexlab{a}})\citenamefont {Corboz}, \citenamefont {Evenbly},
  \citenamefont {Verstraete},\ and\ \citenamefont {Vidal}}]{corboz:2010a}%
  \BibitemOpen
  \bibfield  {author} {\bibinfo {author} {\bibfnamefont {P.}~\bibnamefont
  {Corboz}}, \bibinfo {author} {\bibfnamefont {G.}~\bibnamefont {Evenbly}},
  \bibinfo {author} {\bibfnamefont {F.}~\bibnamefont {Verstraete}}, \ and\
  \bibinfo {author} {\bibfnamefont {G.}~\bibnamefont {Vidal}},\ }\href
  {\doibase 10.1103/PhysRevA.81.010303} {\bibfield  {journal} {\bibinfo
  {journal} {Phys. Rev. A}\ }\textbf {\bibinfo {volume} {81}},\ \bibinfo
  {pages} {010303} (\bibinfo {year} {2010}{\natexlab{a}})}\ 
  \BibitemShut {NoStop}%
\bibitem [{\citenamefont {Corboz}\ \emph
  {et~al.}(2010{\natexlab{b}})\citenamefont {Corboz}, \citenamefont {Or\'us},
  \citenamefont {Bauer},\ and\ \citenamefont {Vidal}}]{corboz:2010b}%
  \BibitemOpen
  \bibfield  {author} {\bibinfo {author} {\bibfnamefont {P.}~\bibnamefont{Corboz}},\
  \bibinfo {author} {\bibfnamefont {R.}~\bibnamefont {Or\'us}},
  \bibinfo {author} {\bibfnamefont {B.}~\bibnamefont {Bauer}}, \ and\ \bibinfo
  {author} {\bibfnamefont {G.}~\bibnamefont {Vidal}},\ }\href {\doibase
  10.1103/PhysRevB.81.165104} {\bibfield  {journal} {\bibinfo  {journal} {Phys.
  Rev. B}\ }\textbf {\bibinfo {volume} {81}},\ \bibinfo {pages} {165104}
  (\bibinfo {year} {2010}{\natexlab{b}})}\ 
  \BibitemShut {NoStop}%
\bibitem [{\citenamefont {Kraus}\ \emph {et~al.}(2010)\citenamefont {Kraus},
  \citenamefont {Schuch}, \citenamefont {Verstraete},\ and\ \citenamefont
  {Cirac}}]{kraus:2010a}%
  \BibitemOpen
  \bibfield  {author} {\bibinfo {author} {\bibfnamefont {C.~V.}\ \bibnamefont
  {Kraus}}, \bibinfo {author} {\bibfnamefont {N.}~\bibnamefont {Schuch}},
  \bibinfo {author} {\bibfnamefont {F.}~\bibnamefont {Verstraete}}, \ and\
  \bibinfo {author} {\bibfnamefont {J.~I.}\ \bibnamefont {Cirac}},\ }\href
  {\doibase 10.1103/PhysRevA.81.052338} {\bibfield  {journal} {\bibinfo
  {journal} {Phys. Rev. A}\ }\textbf {\bibinfo {volume} {81}},\ \bibinfo
  {pages} {052338} (\bibinfo {year} {2010})}\ 
  \BibitemShut {NoStop}%
\bibitem [{\citenamefont {Haegeman}\ \emph {et~al.}(2012)\citenamefont
  {Haegeman}, \citenamefont {Pirvu}, \citenamefont {Weir}, \citenamefont
  {Cirac}, \citenamefont {Osborne}, \citenamefont {Verschelde},\ and\
  \citenamefont {Verstraete}}]{haegeman:2012a}%
  \BibitemOpen
  \bibfield  {author} {\bibinfo {author} {\bibfnamefont {J.}~\bibnamefont
  {Haegeman}}, \bibinfo {author} {\bibfnamefont {B.}~\bibnamefont {Pirvu}},
  \bibinfo {author} {\bibfnamefont {D.~J.}\ \bibnamefont {Weir}}, \bibinfo
  {author} {\bibfnamefont {J.~I.}\ \bibnamefont {Cirac}}, \bibinfo {author}
  {\bibfnamefont {T.~J.}\ \bibnamefont {Osborne}}, \bibinfo {author}
  {\bibfnamefont {H.}~\bibnamefont {Verschelde}}, \ and\ \bibinfo {author}
  {\bibfnamefont {F.}~\bibnamefont {Verstraete}},\ }\href {\doibase
  10.1103/PhysRevB.85.100408} {\bibfield  {journal} {\bibinfo  {journal} {Phys.
  Rev. B}\ }\textbf {\bibinfo {volume} {85}},\ \bibinfo {pages} {100408}
  (\bibinfo {year} {2012})}\ 
  \BibitemShut{NoStop}%
\bibitem [{\citenamefont {Verstraete}\ and\ \citenamefont
  {Cirac}(2004)}]{verstraete:2004a}%
  \BibitemOpen
  \bibfield  {author} {\bibinfo {author} {\bibfnamefont {F.}~\bibnamefont
  {Verstraete}}\ and\ \bibinfo {author} {\bibfnamefont {J.~I.}\ \bibnamefont
  {Cirac}},\ }
  \Eprint {http://arxiv.org/abs/cond-mat/0407066} {arXiv:cond-mat/0407066}\
  (\bibinfo {year} {2004})\BibitemShut {NoStop}%
\bibitem [{\citenamefont {Vidal}(2008)}]{vidal:2006a}%
  \BibitemOpen
  \bibfield  {author} {\bibinfo {author} {\bibfnamefont {G.}~\bibnamefont
  {Vidal}},\ }\href {\doibase 10.1103/PhysRevLett.101.110501} 
  {\bibfield  {journal} {\bibinfo  {journal} {Phys.
  Rev. Lett.}\ }\textbf {\bibinfo {volume} {101}},\ \bibinfo {pages} {110501}
  (\bibinfo {year} {2008})}\ 
  \BibitemShut{NoStop}%
\bibitem [{\citenamefont {Vidal}(2007)}]{vidal:2007a}%
  \BibitemOpen
  \bibfield  {author} {\bibinfo {author} {\bibfnamefont {G.}~\bibnamefont
  {Vidal}},\ }\href {\doibase 10.1103/PhysRevLett.99.220405} 
  {\bibfield  {journal} {\bibinfo  {journal} {Phys.
  Rev. Lett.}\ }\textbf {\bibinfo {volume} {99}},\ \bibinfo {pages} {220405}
  (\bibinfo {year} {2007})}\ 
  \BibitemShut{NoStop}%
\bibitem [{\citenamefont {K\"onig}\ and\ \citenamefont
  {Bilgin}(2010)}]{koenig:2010a}%
  \BibitemOpen
  \bibfield  {author} {\bibinfo {author} {\bibfnamefont {R.}~\bibnamefont
  {K\"onig}}\ and\ \bibinfo {author} {\bibfnamefont {E.}~\bibnamefont
  {Bilgin}},\ }\href {\doibase 10.1103/PhysRevB.82.125118} {\bibfield
  {journal} {\bibinfo  {journal} {Phys. Rev. B}\ }\textbf {\bibinfo {volume}
  {82}},\ \bibinfo {pages} {125118} (\bibinfo {year} {2010})}\BibitemShut
  {NoStop}%
\bibitem [{\citenamefont {Pfeifer}\ \emph {et~al.}(2010)\citenamefont
  {Pfeifer}, \citenamefont {Corboz}, \citenamefont {Buerschaper}, \citenamefont
  {Aguado}, \citenamefont {Troyer},\ and\ \citenamefont
  {Vidal}}]{pfeifer:2010a}%
  \BibitemOpen
  \bibfield  {author} {\bibinfo {author} {\bibfnamefont {R.~N.~C.}\
  \bibnamefont {Pfeifer}}, \bibinfo {author} {\bibfnamefont {P.}~\bibnamefont
  {Corboz}}, \bibinfo {author} {\bibfnamefont {O.}~\bibnamefont {Buerschaper}},
  \bibinfo {author} {\bibfnamefont {M.}~\bibnamefont {Aguado}}, \bibinfo
  {author} {\bibfnamefont {M.}~\bibnamefont {Troyer}}, \ and\ \bibinfo {author}
  {\bibfnamefont {G.}~\bibnamefont {Vidal}},\ }\href {\doibase 10.1103/PhysRevB.82.115126} 
  {\bibfield {journal} {\bibinfo  {journal} {Phys. Rev. B}\ }\textbf {\bibinfo {volume}
  {82}} \bibinfo {pages} {115126} (\bibinfo {year} {2010})}\BibitemShut {NoStop}%
\bibitem [{\citenamefont {Singh}\ \emph {et~al.}(2013)\citenamefont {Singh},
  \citenamefont {Pfeifer}, \citenamefont {Vidal},\ and\ \citenamefont
  {Brennen}}]{singh:2013a}%
  \BibitemOpen
  \bibfield  {author} {\bibinfo {author} {\bibfnamefont {S.}~\bibnamefont
  {Singh}}, \bibinfo {author} {\bibfnamefont {R.~N.~C.}\ \bibnamefont
  {Pfeifer}}, \bibinfo {author} {\bibfnamefont {G.}~\bibnamefont {Vidal}}, \
  and\ \bibinfo {author} {\bibfnamefont {G.~K.}\ \bibnamefont {Brennen}},\
  }
  \href {\doibase 10.1103/PhysRevB.89.075112} {\bibfield
  {journal} {\bibinfo  {journal} {Phys. Rev. B}\ }\textbf {\bibinfo
  {volume} {89}},\ \bibinfo {pages} {075112} (\bibinfo {year} {2014})}\
  \BibitemShut {NoStop}%
\bibitem [{\citenamefont {Poilblanc}\ \emph {et~al.}(2013)\citenamefont
  {Poilblanc}, \citenamefont {Feiguin}, \citenamefont {Troyer}, \citenamefont
  {Ardonne},\ and\ \citenamefont {Bonderson}}]{poilblanc:2013a}%
  \BibitemOpen
  \bibfield  {author} {\bibinfo {author} {\bibfnamefont {D.}~\bibnamefont
  {Poilblanc}}, \bibinfo {author} {\bibfnamefont {A.}~\bibnamefont {Feiguin}},
  \bibinfo {author} {\bibfnamefont {M.}~\bibnamefont {Troyer}}, \bibinfo
  {author} {\bibfnamefont {E.}~\bibnamefont {Ardonne}}, \ and\ \bibinfo
  {author} {\bibfnamefont {P.}~\bibnamefont {Bonderson}},\ }\href {\doibase
  10.1103/PhysRevB.87.085106} {\bibfield  {journal} {\bibinfo  {journal} {Phys.
  Rev. B}\ }\textbf {\bibinfo {volume} {87}},\ \bibinfo {pages} {085106}
  (\bibinfo {year} {2013})}\BibitemShut {NoStop}%
\bibitem [{\citenamefont {Zatloukal}\ \emph {et~al.}(2012)\citenamefont
  {Zatloukal}, \citenamefont {Lehman}, \citenamefont {Singh}, \citenamefont
  {Pachos},\ and\ \citenamefont {Brennen}}]{zatloukal:2012a}%
  \BibitemOpen
  \bibfield  {author} {\bibinfo {author} {\bibfnamefont {V.}~\bibnamefont
  {Zatloukal}}, \bibinfo {author} {\bibfnamefont {L.}~\bibnamefont {Lehman}},
  \bibinfo {author} {\bibfnamefont {S.}~\bibnamefont {Singh}}, \bibinfo
  {author} {\bibfnamefont {J.~K.}\ \bibnamefont {Pachos}}, \ and\ \bibinfo
  {author} {\bibfnamefont {G.~K.}\ \bibnamefont {Brennen}},\ }\href
  {http://arxiv.org/abs/1207.5000} {\bibfield  {journal} {\bibinfo  {journal}
  {{arXiv:1207.5000}}\ } (\bibinfo {year} {2012})}\BibitemShut
  {NoStop}%
\bibitem [{\citenamefont {Zaletel}\ \emph {et~al.}(2013)\citenamefont
  {Zaletel}, \citenamefont {Mong},\ and\ \citenamefont
  {Pollmann}}]{zaletel:2013a}%
  \BibitemOpen
  \bibfield  {author} {\bibinfo {author} {\bibfnamefont {M.~P.}\ \bibnamefont
  {Zaletel}}, \bibinfo {author} {\bibfnamefont {R.~S.~K.}\ \bibnamefont
  {Mong}}, \ and\ \bibinfo {author} {\bibfnamefont {F.}~\bibnamefont
  {Pollmann}},\ }\href {\doibase 10.1103/PhysRevLett.110.236801} {\bibfield
  {journal} {\bibinfo  {journal} {Phys. Rev. Lett.}\ }\textbf {\bibinfo
  {volume} {110}},\ \bibinfo {pages} {236801} (\bibinfo {year}
  {2013})}\BibitemShut {NoStop}%
\bibitem [{\citenamefont {Trebst}\ \emph {et~al.}(2008)\citenamefont {Trebst},
  \citenamefont {Ardonne}, \citenamefont {Feiguin}, \citenamefont {Huse},
  \citenamefont {Ludwig},\ and\ \citenamefont {Troyer}}]{trebst:2008a}%
  \BibitemOpen
  \bibfield  {author} {\bibinfo {author} {\bibfnamefont {S.}~\bibnamefont
  {Trebst}}, \bibinfo {author} {\bibfnamefont {E.}~\bibnamefont {Ardonne}},
  \bibinfo {author} {\bibfnamefont {A.}~\bibnamefont {Feiguin}}, \bibinfo
  {author} {\bibfnamefont {D.~A.}\ \bibnamefont {Huse}}, \bibinfo {author}
  {\bibfnamefont {A.~W.~W.}\ \bibnamefont {Ludwig}}, \ and\ \bibinfo {author}
  {\bibfnamefont {M.}~\bibnamefont {Troyer}},\ }\href {\doibase
  10.1103/PhysRevLett.101.050401} {\bibfield  {journal} {\bibinfo  {journal}
  {Phys. Rev. Lett.}\ }\textbf {\bibinfo {volume} {101}},\ \bibinfo {pages}
  {050401} (\bibinfo {year} {2008})}\BibitemShut {NoStop}%
\bibitem [{\citenamefont {Poilblanc}\ \emph {et~al.}(2011)\citenamefont
  {Poilblanc}, \citenamefont {Ludwig}, \citenamefont {Trebst},\ and\
  \citenamefont {Troyer}}]{poilblanc:2011a}%
  \BibitemOpen
  \bibfield  {author} {\bibinfo {author} {\bibfnamefont {D.}~\bibnamefont
  {Poilblanc}}, \bibinfo {author} {\bibfnamefont {A.~W.~W.}\ \bibnamefont
  {Ludwig}}, \bibinfo {author} {\bibfnamefont {S.}~\bibnamefont {Trebst}}, \
  and\ \bibinfo {author} {\bibfnamefont {M.}~\bibnamefont {Troyer}},\ }\href
  {\doibase 10.1103/PhysRevB.83.134439} {\bibfield  {journal} {\bibinfo
  {journal} {Phys. Rev. B}\ }\textbf {\bibinfo {volume} {83}},\ \bibinfo
  {pages} {134439} (\bibinfo {year} {2011})}\BibitemShut {NoStop}%
\bibitem [{\citenamefont {Poilblanc}\ \emph {et~al.}(2012)\citenamefont
  {Poilblanc}, \citenamefont {Troyer}, \citenamefont {Ardonne},\ and\
  \citenamefont {Bonderson}}]{poilblanc:2012a}%
  \BibitemOpen
  \bibfield  {author} {\bibinfo {author} {\bibfnamefont {D.}~\bibnamefont
  {Poilblanc}}, \bibinfo {author} {\bibfnamefont {M.}~\bibnamefont {Troyer}},
  \bibinfo {author} {\bibfnamefont {E.}~\bibnamefont {Ardonne}}, \ and\
  \bibinfo {author} {\bibfnamefont {P.}~\bibnamefont {Bonderson}},\ }\href
  {\doibase 10.1103/PhysRevLett.108.207201} {\bibfield  {journal} {\bibinfo
  {journal} {Phys. Rev. Lett.}\ }\textbf {\bibinfo {volume} {108}},\ \bibinfo
  {pages} {207201} (\bibinfo {year} {2012})}\BibitemShut {NoStop}%
\bibitem [{\citenamefont {Tran}\ and\ \citenamefont
  {Bonesteel}(2010)}]{tran:2010a}%
  \BibitemOpen
  \bibfield  {author} {\bibinfo {author} {\bibfnamefont {H.}~\bibnamefont
  {Tran}}\ and\ \bibinfo {author} {\bibfnamefont {N.}~\bibnamefont
  {Bonesteel}},\ }\href {\doibase 10.1016/j.commatsci.2010.03.008} {\bibfield
  {journal} {\bibinfo  {journal} {Comput. Mater. Sci.}\ }\textbf
  {\bibinfo {volume} {49}},\ \bibinfo {pages} {S395 } (\bibinfo {year}
  {2010})}\ 
  \BibitemShut {NoStop}%
\bibitem [{\citenamefont {Hastings}\ \emph {et~al.}(2013)\citenamefont
  {Hastings}, \citenamefont {Nayak},\ and\ \citenamefont
  {Wang}}]{hastings:2013a}%
  \BibitemOpen
  \bibfield  {author} {\bibinfo {author} {\bibfnamefont {M.~B.}\ \bibnamefont
  {Hastings}}, \bibinfo {author} {\bibfnamefont {C.}~\bibnamefont {Nayak}}, \
  and\ \bibinfo {author} {\bibfnamefont {Z.}~\bibnamefont {Wang}},\ }\href
  {\doibase 10.1103/PhysRevB.87.165421} {\bibfield  {journal} {\bibinfo
  {journal} {Phys. Rev. B}\ }\textbf {\bibinfo {volume} {87}},\ \bibinfo
  {pages} {165421} (\bibinfo {year} {2013})}\BibitemShut {NoStop}%
\bibitem [{\citenamefont {Bonderson}(2007)}]{Bond07}%
  \BibitemOpen
  \bibfield  {author} {\bibinfo {author} {\bibfnamefont {P.~H.}\ \bibnamefont
  {Bonderson}}\ }\href@noop {} {\bibfield  {journal} {\bibinfo  {journal} {PhD thesis, California
	Institute of Technology}\ }  (\bibinfo {year} {2007})}\BibitemShut {NoStop}%
\bibitem{EvoMPS}
  A.~Milsted, evoMPS source code (2013),
  \url{https://github.com/amilsted/evoMPS}.
\bibitem [{\citenamefont {Fannes}\ \emph {et~al.}(1992)\citenamefont {Fannes},
  \citenamefont {Nachtergaele},\ and\ \citenamefont {Werner}}]{Werner1992}%
  \BibitemOpen
  \bibfield  {author} {\bibinfo {author} {\bibfnamefont {M.}~\bibnamefont
  {Fannes}}, \bibinfo {author} {\bibfnamefont {B.}~\bibnamefont
  {Nachtergaele}}, \ and\ \bibinfo {author} {\bibfnamefont {R.~F.}\
  \bibnamefont {Werner}},\ }\href
  {http://link.springer.com/article/10.1007/BF02099178} {\bibfield  {journal}
  {\bibinfo  {journal} {Commun. Math. Phys.}\ }\textbf
  {\bibinfo {volume} {144}},\ \bibinfo {pages} {443} (\bibinfo {year}
  {1992})}\BibitemShut {NoStop}%
\bibitem [{\citenamefont {Verstraete}\ \emph {et~al.}(2008)\citenamefont
  {Verstraete}, \citenamefont {Murg},\ and\ \citenamefont
  {Cirac}}]{Verstraete2008}%
  \BibitemOpen
  \bibfield  {author} {\bibinfo {author} {\bibfnamefont {F.}~\bibnamefont
  {Verstraete}}, \bibinfo {author} {\bibfnamefont {V.}~\bibnamefont {Murg}}, \
  and\ \bibinfo {author} {\bibfnamefont {J.}~\bibnamefont {Cirac}},\ }\href
  {http://www.tandfonline.com/doi/full/10.1080/14789940801912366} {\bibfield
  {journal} {\bibinfo  {journal} {Adv. Phys.}\ }\textbf {\bibinfo
  {volume} {57}},\ \bibinfo {pages} {143} (\bibinfo {year} {2008})}\BibitemShut
  {NoStop}%

\bibitem{milsted_phi4}%
  \BibitemOpen
  \bibfield  {author} {\bibinfo {author} {\bibfnamefont {A.}~\bibnamefont {Milsted}},\
  \bibinfo {author} {\bibfnamefont {J.}~\bibnamefont
  {Haegeman}}, \bibinfo {author} {\bibfnamefont {T.~J.}~\bibnamefont {Osborne}},}
  \href {\doibase 10.1103/PhysRevD.88.085030} 
  {\bibfield {journal} {\bibinfo  {journal} {Phys. Rev. D}\ }\textbf {\bibinfo {volume}
  {88}} \bibinfo {pages} {085030} (\bibinfo {year} {2013})}\BibitemShut
{NoStop}%
\bibitem{fes}
  \BibitemOpen
  \bibfield  {author} {\bibinfo {author} {\bibfnamefont {L.}~\bibnamefont
  {Tagliacozzo}}, \bibinfo {author} {\bibfnamefont {T.}~\bibnamefont {de Oliveira}}, \
  \bibinfo {author} {\bibfnamefont {S.}~\bibnamefont {Iblisdir}}, \
  and\ \bibinfo {author} {\bibfnamefont {J.}~\bibnamefont {Latorre}},\ }\href
  {http://link.aps.org/doi/10.1103/PhysRevB.78.024410} {\bibfield
  {journal} {\bibinfo  {journal} {Phys. Rev. B}\ }\textbf {\bibinfo
  {volume} {78}},\ \bibinfo {pages} {024410} (\bibinfo {year} {2008})}\BibitemShut
  {NoStop}%
\bibitem{conformal_fes}
  \BibitemOpen
  \bibfield  {author} {\bibinfo {author} {\bibfnamefont {V.}~\bibnamefont
  {Stojevic}}\, \bibinfo {author} {\bibfnamefont {J.}~\bibnamefont {Haegeman}}, \
  \bibinfo {author} {\bibfnamefont {I.~P.}~\bibnamefont {McCulloch}}, \
  \bibinfo {author} {\bibfnamefont {L.}~\bibnamefont{Tagliacozzo}} \
  and\ \bibinfo {author} {\bibfnamefont {F.}~\bibnamefont
  {Verstraete}},\ }\href@noop {} 
  \Eprint {http://arxiv.org/abs/1401.7654} {arXiv:1401.7654} \BibitemShut
  {NoStop}%
\bibitem [{\citenamefont {Calabrese}\ and\ \citenamefont
  {Cardy}(2004)}]{Cardy2004}%
  \BibitemOpen
  \bibfield  {author} {\bibinfo {author} {\bibfnamefont {P.}~\bibnamefont
  {Calabrese}}\ and\ \bibinfo {author} {\bibfnamefont {J.}~\bibnamefont
  {Cardy}},\ }\href {http://iopscience.iop.org/1742-5468/2004/06/P06002}
  {\bibfield  {journal} {\bibinfo  {journal} {J. Stat. Mech.
  }\ }\textbf {\bibinfo {volume} {2004}},\ \bibinfo
  {pages} {P06002} (\bibinfo {year} {2004})}\BibitemShut {NoStop}%
\bibitem [{\citenamefont {Birman}\ and\ \citenamefont {Wenzl}(1989)}]{BiWe89}%
  \BibitemOpen
  \bibfield  {author} {\bibinfo {author} {\bibfnamefont {J.~S.}\ \bibnamefont
  {Birman}}\ and\ \bibinfo {author} {\bibfnamefont {H.}~\bibnamefont {Wenzl}},\
  }\href@noop {} {\bibfield  {journal} {\bibinfo  {journal} {Trans. AMS}\
  }\textbf {\bibinfo {volume} {313}},\ \bibinfo {pages} {249} (\bibinfo {year}
  {1989})}\BibitemShut {NoStop}%
\bibitem [{\citenamefont {Murakami}(1987)}]{Mura87}%
  \BibitemOpen
  \bibfield  {author} {\bibinfo {author} {\bibfnamefont {J.}~\bibnamefont
  {Murakami}},\ }\href@noop {} {\bibfield  {journal} {\bibinfo  {journal}
  {Osaka J. Math.}\ }\textbf {\bibinfo {volume} {24}},\ \bibinfo {pages} {745}
  (\bibinfo {year} {1987})}\BibitemShut {NoStop}%
\bibitem [{\citenamefont {Temperley}\ and\ \citenamefont
  {Lieb}(1971)}]{TeLi71}%
  \BibitemOpen
  \bibfield  {author} {\bibinfo {author} {\bibfnamefont {H.~N.~V.}\
  \bibnamefont {Temperley}}\ and\ \bibinfo {author} {\bibfnamefont {E.~H.}\
  \bibnamefont {Lieb}},\ }\href@noop {} {\bibfield  {journal} {\bibinfo
  {journal} {Proc. R. Soc. Lond. A}\ }\textbf {\bibinfo {volume} {332}},\
  \bibinfo {pages} {251} (\bibinfo {year} {1971})}\BibitemShut {NoStop}%
\bibitem [{\citenamefont {Owczarek}\ and\ \citenamefont
  {Baxter}(1987)}]{OwBa87}%
  \BibitemOpen
  \bibfield  {author} {\bibinfo {author} {\bibfnamefont {A.~L.}\ \bibnamefont
  {Owczarek}}\ and\ \bibinfo {author} {\bibfnamefont {R.~J.}\ \bibnamefont
  {Baxter}},\ }\href@noop {} {\bibfield  {journal} {\bibinfo  {journal} {J.
  Stat. Phys.}\ }\textbf {\bibinfo {volume} {49}},\ \bibinfo {pages} {1093}
  (\bibinfo {year} {1987})}\BibitemShut {NoStop}%
\bibitem [{\citenamefont {Aufgebauer}\ and\ \citenamefont
  {Kl{\"u}mper}(2010)}]{AuKl10}%
  \BibitemOpen
  \bibfield  {author} {\bibinfo {author} {\bibfnamefont {B.}~\bibnamefont
  {Aufgebauer}}\ and\ \bibinfo {author} {\bibfnamefont {A.}~\bibnamefont
  {Kl{\"u}mper}},\ } 
  \href{\doibase 10.1088/1742-5468/2010/05/P05018}{\bibfield  {journal} {\bibinfo  {journal}
  {J. Stat. Mech.}\ ,\ \bibinfo {pages} {P05018}} (\bibinfo {year} {2010})}\
  \BibitemShut
  {NoStop}%
\bibitem [{\citenamefont {Cloizeaux}\ and\ \citenamefont
  {Gaudin}(1966)}]{ClGa66}%
  \BibitemOpen
  \bibfield  {author} {\bibinfo {author} {\bibfnamefont {J.~D.}\ \bibnamefont
  {Cloizeaux}}\ and\ \bibinfo {author} {\bibfnamefont {M.}~\bibnamefont
  {Gaudin}},\ }\href@noop {} {\bibfield  {journal} {\bibinfo  {journal} {J.
  Math. Phys.}\ }\textbf {\bibinfo {volume} {7}},\ \bibinfo {pages} {1384}
  (\bibinfo {year} {1966})}\BibitemShut {NoStop}%
\bibitem [{\citenamefont {Fateev}\ and\ \citenamefont
  {Zamolodchikov}(1982)}]{FaZa82}%
  \BibitemOpen
  \bibfield  {author} {\bibinfo {author} {\bibfnamefont {V.~A.}\ \bibnamefont
  {Fateev}}\ and\ \bibinfo {author} {\bibfnamefont {A.~B.}\ \bibnamefont
  {Zamolodchikov}},\ }\href{\doibase 10.1016/0375-9601(82)90736-8}
  {\bibfield  {journal} {\bibinfo  {journal}
  {Phys. Lett. A}\ }\textbf {\bibinfo {volume} {92}},\ \bibinfo {pages} {37}
  (\bibinfo {year} {1982})}\BibitemShut {NoStop}%
\bibitem [{\citenamefont {Albertini}(1994)}]{Albe94}%
  \BibitemOpen
  \bibfield  {author} {\bibinfo {author} {\bibfnamefont {G.}~\bibnamefont
  {Albertini}},\ }\href@noop {} {\bibfield  {journal} {\bibinfo  {journal}
  {Int. J. Mod. Phys. A}\ }\textbf {\bibinfo {volume} {9}},\ \bibinfo {pages}
  {4921} (\bibinfo {year} {1994})},\
  \Eprint{http://arxiv.org/abs/hep-th/9310133} {arXiv:hep-th/9310133} 
  \BibitemShut {NoStop}%
\bibitem [{\citenamefont {Bazhanov}\ and\ \citenamefont
  {Stroganov}(1990)}]{BaSt90}%
  \BibitemOpen
  \bibfield  {author} {\bibinfo {author} {\bibfnamefont {V.~V.}\ \bibnamefont
  {Bazhanov}}\ and\ \bibinfo {author} {\bibfnamefont {Yu.~G.}\ \bibnamefont
  {Stroganov}},\ }\href@noop {} {\bibfield  {journal} {\bibinfo  {journal}
  {J. Stat. Phys.}\ }\textbf {\bibinfo {volume} {59}},\ \bibinfo {pages} {799}
  (\bibinfo {year} {1990})}\BibitemShut {NoStop}%
\bibitem{FiFF14}%
  \BibitemOpen
  \bibfield  {author} {\bibinfo {author} {\bibfnamefont {P.~E.~}\ \bibnamefont
  {Finch}}\ and\ \bibinfo {author} {\bibfnamefont {M.}\ \bibnamefont
  {Flohr}}\ and\ \bibinfo {author} {\bibfnamefont {H.}\ \bibnamefont
  {Frahm}},\ }\href@noop {} \Eprint{http://arxiv.org/abs/1408.1282} {arXiv:1408.1282} 
  \BibitemShut {NoStop}%
\end{thebibliography}

%

\end{document}